
\magnification 1200
\parindent=16pt
\baselineskip=24pt
  
  \def\avt#1{\langle#1\rangle}
  \vskip 2.truecm\noindent
  \line{}
  \line{}
  \centerline{\bf Self-affine Asperity Model for earthquakes}
  \vskip 1.4truecm\noindent
  \centerline{V. De Rubeis$^{1}$, R. Hallgass$^{2}$, V. Loreto$^{2}$, }
  \medskip
  \centerline{G. Paladin$^3$, L. Pietronero$^2$ and P. Tosi$^{1}$}
  \vskip .6truecm
  \centerline{\it $^1$Istituto Nazionale di Geofisica}
  \centerline{\it  Via di Vigna Murata 605 I-00143 Roma, Italy}
  \vskip .4truecm
  \centerline{\it $^2$Dipartimento di Fisica, Universit\`a di Roma
  'La Sapienza'}
  \centerline{\it P.le A.Moro 2 I-00185 Roma, Italy}
  \vskip .4truecm
  \centerline{\it $^3$Dipartimento di Fisica,  Universit\`a dell'Aquila}
  \centerline{\it  Via Vetoio I-67100 Coppito, L'Aquila, Italy}
  \vskip 1.6truecm
  \centerline{ABSTRACT}
  A model  for fault dynamics consisting of two rough
  and rigid brownian profiles that slide one over the other is introduced.
  An earthquake occurs when there is an intersection between the
  two profiles. The energy release is proportional to the overlap interval.
  Our  model  exhibits  some specific features which follow from the
  fractal geometry of the fault:
  (1) non-universality of the exponent of the Gutenberg-Richter law
  for the magnitude distribution;
  (2) presence of local stress accumulation before
  a large seismic event; (3) non-trivial space-time clustering
  of the epicenters. These properties are in good agreement with
  various observations and lead to specific predictions that can be
  experimentally tested.
  \vskip .4truecm
  \noindent
   \noindent
  PACS NUMBERS:  91.30.Px, 05.40.+j
  \vfill\eject
  \noindent

  Many forms of scaling invariance appear in seismic phenomena: the  celebrated
  Gutenberg-Richter law for the magnitude distribution  [1],
  the Omori law for the time correlations of aftershocks
   [2], space-time  clustering of the epicenters  [3] are
  a common mark of the earthquake statistics.
  Unfortunately, the complexity  of  modelling the motion of a fault
  system, even in rather well controlled situation such as the San
  Andreas fault in California, is a highly difficult task and it is
  still controversial what is the correct theoretical framework at
  the very origin of scaling laws. It is thus important to
  individuate models as simple as possible that are able to exhibit
  the main  qualitative features of the fault dynamics.
  Their physical relevance stems from the specific predictions on the
  {\it real} seismic activity which might be verified from experimental data.

  One of the first attempt in this direction is due to Burridge and
  Knopoff  [4] who introduced a stick-slip model  of coupled
  oscillators to mimic the interaction of two fault surfaces.  In practice
  one considers blocks on a rough support connected to one  another
  by springs.
  They are also connected by other springs to a driver which moves
  at a very low constant velocity. The blocks stick until the
  spring forces overwhelms the static friction
  and then one or more blocks slide, releasing an `earthquake' energy
  proportional to the sum of the displacements.
  A numerical integration of the Newton equations for
  a one-dimensional  chain with a large number of homogeneous blocks
  have  been showed to exhibit the Gutenberg-Richter law [5]
   (see also [6] for the connection with the chaotic behaviour
   of the system).
  Moreover, it has been proposed that the qualitative aspects of
  earthquakes (and of Burridge and Knopoff models) are captured by the
  so-called
  Sandpile models, which represent the paradigm of a large class of
  Self-Organized Critical (SOC) systems [7], where the scaling is spontaneously
  generated by the dynamics. In fact, there is a whole generation of SOC models
  to explain the scale invariant properties of earthquakes [8,9].
  These type of models suggest however that there is no stress accumulation
  before a big earthquake and the exponent of the Gutenberg-Richter law is
  expected (with some exceptions  [10])  to be universal. In addition the
  space-time distribution of the epicenters has no clear relation with
  the experiments where non-trivial clustering and correlations are present.

  In order to go beyond these limitations we propose here an alternative
  approach where the critical behavior is
  not self-organized  but stems from the fractal geometry of the fault
that is supposed to arise as a consequence of geological processes on very long
  time scales with respect to the seismic dynamics.
  Looking at the system on the time scale of human records the fault structure
  can be considered assigned and just slightly modified by earthquakes.

  Many authors pointed out that natural rock surfaces
  are represented by fractional brownian surfaces over a wide scale range
  [11] and that also the topographic traces of the fault
  surfaces exhibit scale invariance [12].
  A fault can thus be regarded as a
  statistically self-affine profile $h(x)$, whose height scales as
  $|h(x+\ell)-h(x)|\sim \ell^{H}$. In $d=2$, such a profile
  $h(x)$ can be generated by fractional brownian motion with exponent
  $H$ and in $d=3$ by the standard generalization given by
  brownian reliefs [13,14].
  The exponent $0\le H \le 1$ controls the roughness of the fault
  where the standard random walk profile corresponds to $H=1/2$,
  and a differentiable curve corresponds to $H=1$.
  The fractal dimension of the profile is well known to be $D_F=d-H$.

  Let us now introduce the self-affine asperity model (SAM)
  that is, in a certain sense, the limit of infinite rigidity
  of the Burridge-Knopoff models and it represents an alternative limit
  with respect to the SOC models.
  The  model is defined by the following dynamical rules:
  {\bf (i)} We consider two profiles, say $h_1(x)$ and $h_2(x)$,
  on parallel supports of length $L$ at infinite distance.
  The initial condition is obtained by putting them in contact in
  the point where the height difference is minimal so that
  $h_1-h_2 \ge 0 \, , \ \forall \, x \in [0,L]$ (see Fig. 1a);
  {\bf (ii)} The successive evolution is obtained by drifting
  a profile in a parallel way with respect to the other one,
  at a constant speed $v$, so that  $h_1(x;t)=h_1(x-v \,t)$;
  {\bf (iii)} At each time step $t$, one controls whether there
  are new contact points between the profiles, i.e.
  whether $h_1(x;t)-h_2(x) < 0$ for some $x$ value.
  An intersection represents a single seismic event
  and starts with the collision of two {\it asperities} of the profiles.
  The energy released is assumed to be proportional to
  the extension of the overlap between the two asperities
  in contact, see Fig. 1b;
  {\bf (iv)} We do not allow the developing of new earthquakes
  in a region where a seismic event is already taking place .

  With these rules, the motion of the two profiles simulate the
  slipping of the two walls of a single fault. The points of collision are
  the points of the fault where the morphology prevents the free
  slip: these are the points where there is an accumulation of stress
  and, consequently, a raise of pressure.
  When the local pressure exceeds a certain threshold, it happens
  a breaking, an earthquake, which allows to relax the stress and
  redistribute the energy, previously accumulated, all around.
  Rule (iii) of the SAM  stems from  the fact that the magnitude of a
  real  earthquake is proportional to the  log of the seismic moment
  $M_0$, which, on its turn, is proportional to the average displacement
  of the fault according to the standard geophysical definition.

  For sake of simplicity, in the SAM, there is no real breaking of the
  profiles as a consequence of an earthquake and  the profiles maintain
  their structures after a crash.
  It is possible to introduce a more realistic breaking mechanism
  where there is  also a modification  of the asperity form
  after an earthquake. However, we have verified that the main
  qualitative features remain unchanged. So we are in the opposite
  perspective than SOC models. In our case
  the earthquake dynamics has no effect on the structure of the profile.
  Realistic situation could well correspond to intermediate cases of course.

  It is worth to stress that the SAM exhibits a strong
  non-locality since  a collision in a point $x$, at the time $t$ can
  trigger, at later time, a subsequent event also very far away.
  One of the main advantage of the SAM consists in the possibility of
  deriving various analytic results using the properties of brownian profiles.
  The most impressive characteristic of the earthquake statistics
  is the Gutenberg-Richter law. It states that the probability
  $P(E)\;dE$ that an earthquake releases an energy in the interval
  $[ \, E \, , \, E+dE \, ]$ scales according to a power law
  $  P(E) \sim E^{-\beta-1}$
  with an exponent $\beta$  of order of the unity [10].
  It is a controversial issue whether $\beta$ is universal or varies
  in a narrow range according to the characteristics of the fault system.

  In the framework of our model it is possible to relate the value of
  the exponent $\beta$ to the geometrical properties of the faults.
  In particular it can be showed that:
  $$
  \beta=1-{H\over (d-1)}={D_F-1\over d-1}.
  \eqno(1)
  $$
  This relation accounts for the direct dependence of the $\beta$-exponent
  on the roughness of the faults $H$.

  In order to derive (1), consider the profile
  $h_1(x;t)-h_2(x)$, which,
 being given by the difference of two
  brownian profiles is, on its turn, a brownian profile at any time $t$.
The statistics of the intersections between the two profiles
is then given by the statistics of the intersections of the brownian profile
 difference with a straight line along the temporal axis.
 Due to the invariance under temporal shifts of the profile, we
 can assume that the statistics of the intersections obtained at any time
with a profile difference  is given by the statistics of the intersections of
an infinite profile with a zero level straight line.

  In this perspective, a seismic event releases an energy proportional
  to the interval between two sub-sequent intersections between a
  brownian profile and  the zero level straight line.
  It is well known that the set obtained by the intersection
  between a fractional brownian profile or relief of dimension $d-H$
  embedded in a $d$-dimensional space  and a hyper-surface  of
  dimensionality $d-1$, is a fractal with dimension given by the law
  of addition of the codimension  [13], $(d-H)+(d-1)-d$,
  so that the number of intersections in a hyper-surface of volume
  $E \sim L^{d-1}$ scales as:  $  N(L) \sim L^{d-1-H}. $
  Now, if we identify the energy released from an earthquake
  with the size $E$ of an intersection, we can determine the exponent
  $\beta$ by consistency requirements.
  In our case the probability $P(E)$
   is given by the probability of finding
  an intersection, of size $E$,
  between a $d$-dimensional surface and a $(d-1)$-dimensional
  hyperplane.
  As a consequence of the geometric properties of the model,
  $P(E)$ follows a  power law.
  Let us consider the average value of the intersection size:
  $$
  {\avt{E}}\equiv \int_{0}^{L^{d-1}} P(E) \, E \, dE \sim
  L^{(d-1)(1-\beta)}.
  \eqno(2)
  $$
  While the typical length of a $d-1$-dimensional
  interval is the total length $L^{d-1}$ of the support divided by
  the number of intersection $N(L)$  so that:
  $  {\avt{E}}=L^{d-1} / N(L) \sim L^{H}.$
  Therefore, one gets $H=(d-1)(1-\beta)$
  and $\beta=1-{H \over d-1}$ that leads to eq.(1).

  It is interesting to notice that the value $\beta=1$ is an upper bound
  reached when the roughness of the fault is maximal ($H=0$).
  Moreover $\beta=1$ is also recovered  for all $H$-values in the mean
  field limit $d \to \infty$, while at $d=3$, $\beta$ can vary in the
  range $[0.5 \, , \,  1]$.

  We have performed numerical simulations by considering two brownian profiles,
  one of which at rest and composed by $10^4$ points and the other,
  slipping
  over the first one, composed by $2 \cdot 10^4$ points. In this way
  each realization of the dynamics lasts a time $T=10^4$.
  The probability distribution of earthquakes has been obtained by averaging
over many
  realizations of the dynamics.
  Fig. 2 shows the numerical results
  in the case of $H=0.5$ and $d=2$. The exponent of the power law
  in this case is $\beta=0.5$ in good agreement with our theoretical
prediction.
  The Gutenberg Richter law is obtained by the cumulative distribution of the
  frequency of earthquakes, i.e. the integral of the distribution showed in
figure.

  Another interesting feature that can be studied in the framework
  of the SAM is the phenomenology of the space-time correlations
  of earthquakes. In particular we will focus on the problem of the spatial
  clustering of epicenters [15] and we refer to  [16] for a more
  exhaustive treatment of this point, including the analysis of the
  correlation functions and the temporal fractal distribution
  of epicenters.
  In our model the space location of an epicenter is defined
  in correspondence of the first point of contact of the two profiles.
  Numerical simulations, performed on the SAM in the cases with $H=0.3$,
  $H=0.5$ and $H=0.7$seem to provide a clear evidence, see fig. 3,
  of a spatial clustering of the epicenters on a set
  with a fractal dimension smaller than $1$ ($D_F \simeq 0.78$
  in the case with $H=0.5$).
  However, this result is a non-trivial finite size effect,
   since the set of epicenters tends to be compact.
   In fact it can be proved, for $H=0.5$,
   that the fractal dimension $D_F(L)$ of the epicenters set in
   a fault of a linear size $L$  is:
  $$
  D_F(L) \simeq 1- \gamma{\log \log L \over \log L }, \ {\rm for \ large \  L}
  \eqno(3)
  $$
  Let us, indeed, consider two brownian profiles of length $L$ as in Fig. 1a.
  The distance $h_{0}(L)$ between the barycentre of the two profiles
  can be obtained from the Iterated Logarithm Theorem  [17] which states
  that, for a partial sum $S_{n}=\sum_{i=1}^{n}x_i$ of identically distributed
  random variables $x_i$ with $<x_i>=0$ and $<x_i^2>=1 \,
  \forall i \in 1,..,n$, it holds:
  $$
  P\left( \limsup_{n \rightarrow \infty}{S_n \over \sqrt{2n \log \log n}}=
  1 \right)=1.
  \eqno(4)
  $$
  That means that the maximum $M(L)$ of a brownian
  profile scales as
  $
  M(L) \sim \sqrt{2 L \log \log L}.
  $

  Now, the distance $h_0 (L)$ is given exactly by the maximum value of a
  brownian profile obtained by the difference of two brownian profiles,
  that is  $  h_0 (L) \sim M(L).$
  On the basis of this result, it is possible to estimate how the number
  of epicenters scales  as a function of $L$.
  Considering
  the configuration where two brownian profiles are $h_0 (L)$ apart,
  the number of points of the lower profile at a certain
  height $h$ with respect to its barycentre, is:
  $$
   N_{down} \sim \sqrt{L} \exp{-\left( h^2 \over 2 \eta \, L \right)}
  \eqno(5)
  $$
  where $\eta$ is a constant depending on the value of $<x_i^2>$ [18].
  We have now to integrate over all the possible values of $h$ that correspond
  to the heights at which there could be an intersection of the two
  profiles in order to obtain the number of events ($N_{ep}$).
  The two integration extremes are given by the maximum value of the lower
  profile and the minimum value of the upper one, that is:
  $$
  N_{ep}(L) \sim \sqrt{L}
  \int_{h_0 - \sqrt{2L \log \log L}}^{\sqrt{2L \log \log L}}
  \exp{- \left( h^2 \over 2 \eta \, L \right) } \, dh
 \sim {L \over (\log L)^{\gamma}} \sqrt{\log \log L},
  \eqno(6)
  $$
 where $\gamma=\alpha / \eta$ and $\alpha$
  is an intermediate value between $\sqrt{2}-1$ and $1$.
  Using the mass-length definition of fractal dimension,
  $ D_F(L)= \log N_{ep}(L) / \log L $,
  relation (3) is proved.
   The asymptotic value $D_F =1$ is reached very slowly at increasing $L$
  and it cannot be detected but by huge simulations.
  We have checked the validity of (5) for profiles
   with a linear size $L$ varying in the range $10^2 \, - \, 10^6$.
  Work is in progress to extend our  results to the case of a generic roughness
  index $H$ [18].

  In summary, we have proposed a model of earthquakes where
  the critical behavior is generated by a pre-existent fractal geometry
  of the fault. The statistics of earthquakes is thus related to
  the roughness of the fault via the scaling relation (1)
  between critical indices.
  This result suggests that the younger the fault system,
  the larger the $\beta$ exponent, since the roughness
  of a fault is expected to decrease in geological times.
  The exponent $\beta$ therefore is non-universal.
 The model exhibits complex space-time correlations between epicenters:
 from the temporal point of view, there exists a fractal
clusterization [16], although the spatial
fractal distribution of the epicenters
 turns out to be a finite size effect very difficult to be detected from
  data analysis. Our model provides a possible explanation for the
  highly irregular and non random distribution of epicenters that is
  experimentally observed.
Moreover, the accumulation of pressure
is at the very origin of large seismic events in the SAM.
  The presence of such an effect could be tested
 also in real situations e.g. by piezo-electric measurements.
  \medskip
  We are grateful for interesting discussions to E. Caglioti,  O. Mazzella
   and  R. Scarpa.
  \vfill\eject
  \centerline {\bf Figure captions}
  \bigskip
  \item{Fig. 1} (a) Example of two brownian profiles modelling the fault
surfaces.
  (b) Sketch for the definition of the energy released during an earthquake:
  it is assumed to be proportional to the overlap interval
  of the two fault surfaces during the slip.
  \medskip
  \item{Fig. 2} Probability density of the earthquakes releasing an energy $E$
vs. $E$
  for roughness index
  $H=0.5$.
  \medskip
  \item{Fig. 3} Box-counting analysis of the spatial distribution of epicenters
  for roughness index
  $H=0.5$ in a system with linear dimension $L=10^4$. The distribution
apparently
  shows a fractal dimension $D_F=0.78$.
  \vfill\eject
   \vskip 0.4truecm
  \noindent
  \centerline {\bf REFERENCES}
  \vskip 0.4truecm
  \item{[1] } Gutenberg B. and Richter C.F.  1956 Ann. Geophys. {\bf 9}, 1.
  \item{[2] } Omori F. 1894,  Rep. Earth. Inv. Comm., 2 , 103.
  \item{[3] } Kagan Y.Y. and Knopoff L. 1980, Geophys. J. R. Astron. Soc.,
  {\bf 62}, 303.
  \item{[4] } Burridge R. and Knopoff L. 1967, Bull. Seismol. Soc. Am. {\bf
57}, 341.
  \item{[5]} Carlson J.M. and Langer J.S. 1989, Phys. Rev. Lett. {\bf 62},
2632;
  Phys. Rev. {\bf A40}, 6470
  \item{[6]} Crisanti A., Jensen M.H., Vulpiani A. and Paladin G. 1993
  Phys. Rev. {\bf A46} R7363
  \item{[7]} Bak P., Tang C. and Wiesenfeld K. 1987, Phys. Rev. Lett. {\bf 59},
381;
  1988 Phys. Rev. A {\bf 38}, 364.
  \item{[8]} Bak P. and Tang C. 1989, J. Geophys. Res. {\bf 94}, 15635.
  \item{[9]} Ito K. and Matsuzaki M. 1990, J. Geophys. Res. {\bf 95}, 6853.
  \item{[10]} Olami Z., Feder H. J. S. and Christensen 1992, Phys. Rev. Lett.
{\bf 68} 1244;
  Christensen K. and Olami Z. 1992, J. Geophys. Res. {\bf 97}, 8729.
  \item{[11]} Brown S.R. and Scholz C.H. 1985, J. Geophys. Res., {\bf
90}, 12575; Wu R.S. and Aki K. 1985, PAGEOPH {\bf 123}, 805.
  \item{[12]} Power W., Tullis T., Brown S., Boitnott G. and Scholz C.H.
1987, Geophys. Res. Lett. {\bf 14}, 29.
  \item{[13]} Mandelbrot B. 1983, The fractal geometry of nature,
  Freeman and Co., New York, pp 256-258.
  \item{[14]} Turcotte D. L. 1992, Fractals and chaos in geology and
  geophysics, Cambridge Un. Press, Cambridge.
\item{[15]}
 Smalley R.F. Jr., Chatelain J.L., Turcotte D.L. and Pr\'evot R.: 1987 Bull.
Seis. Soc.
 Am. {\bf 77}, 1378; Hirata T.: 1989 J. Geophys. Res. {\bf 94}, 7507;
 Kagan Y.Y. and Jackson D.D.: 1991, Geophys. J. Int. {\bf 104}, 117;
 Main I. G. 1992, Geophys. J. Int. {\bf 111}, 531;
De Rubeis V., Dimitriu P., Papadimitriou E. and Tosi P. 1993, Geophys. Res.
Lett.
{\bf 20}, 1911.
\item{[16]} V. De Rubeis, R. Hallgass, V. Loreto, L. Pietronero and P.Tosi:
1995
in preparation.
 \item{[17]} Grimmett G. R., Stirzaker D. R. 1992, Probability and Random
Processes
  second edition, Oxford Science Publications, New York.
\item{[18]}R. Hallgass, V. Loreto, O. Mazzella and G. Paladin: 1995 in
preparation.
  \bye